\documentstyle[preprint,aps,graphicx,eqsecnum]{revtex}

\begin{document}
\draft
\title{Phenomenological analysis of lepton and quark Yukawa couplings
in SO(10) two Higgs model}

\author{Koichi MATSUDA}
\address{
Graduate school of Science, 
Osaka University, Toyonaka, Osaka 560-0043, Japan}

\maketitle

\begin{abstract}
We investigate a model that the Yukawa coupling form is constructed 
by two kinds of matrix (\(M_0\) and \(M_1\)). 
For example, in the \(SO(10)\) GUT model, \(M_0\) and \(M_1\) are 
Yukawa couplings generated 
by the \(10\) and \(126\) Higgs scalars.
We study how this model can give the observed mass and mixings of 
quarks and leptons. 
Parameter fitting is fully scanned by assuming all the input data to be 
normally distributed around the center value.
\end{abstract}

\section{Introduction}
The grand unification theory (GUT) is very attractive as an unified 
description of the fundamental forces in the nature.  
However, in order to reproduce the observed quark and charged-lepton masses 
and mixings, a lot of Yukawa couplings are usually brought into the model.  
We think that the nature is simple. 
So it is the very crucial problem to know the minimum number 
of Yukawa couplings which can give the observed fermion mass spectra and mixings. 
However, if the quark and lepton Yukawa couplings are composed by only one matrix 
\begin{equation}
M_u = c_u M_0 , \quad  M_d = c_d M_0 , \quad M_e = c_e M_0,
\end{equation}
the CKM Matrix must be diagonalized, these model is obviously ruled out for the 
description of realistic quark and lepton mass spectra.
Therefore, at the unification scale $\mu=\Lambda_X$,
we assume the Yukawa coupling of up quark, down quark and charged-lepton 
(\(M_u^0\), \(M_d^0\) and \(M_e^0\)) are composed by two matrices,
\begin{equation}
M_f^0= c_{f0} M_0^0+ c_{f1} M_1^0. 
\qquad (f=u,\ d,\ e
)\label{eq90101}
\end{equation}
Here, \(c_{f0}\) and \(c_{f1}\) are real numbers which can be associated with 
the vacuum expectation values (VEV).
For example, in the SO(10) GUT model with one 10 and one 126 Higgs scalars, 
the Yukawa couplings of quarks and charged 
leptons are expressed in the following forms \cite{twohiggs1}\cite{twohiggs2}:
\begin{eqnarray}
M_u^0&=&c_0M_0^0+c_1M_1^0,~ M_d^0=M_0^0+M_1^0,~ M_e^0=M_0^0- 3 M_1^0
. \label{eq110501}
\end{eqnarray}
where $M_0^0$ and $M_1^0$ are symmetric Yukawa couplings.
In the previous paper \cite{twohiggs1}, 
eliminating $M_0^0$ and $M_1^0$ from Eq.(\ref{eq90101}), 
we obtain the relation
\begin{equation}
M_e^0 = c_u M_u^0 + c_d M_d^0 \equiv c_u (M_u^0 + \kappa M_d^0), \label{eq82515}
\end{equation}
where
\begin{equation}
c_d=\frac{c_{u0}c_{e1} - c_{e0} c_{u1}}{c_{u0}c_{d1} - c_{d0} c_{u1}}
\qquad \mbox{ and } \qquad 
c_u=\frac{c_{e0} c_{d1} - c_{d0} c_{e1}}{c_{u0}c_{d1} - c_{d0} c_{u1}}.
\end{equation}
These relations are realized at the GUT scale, 
but each value of the Yukawa couplings is given by the experiment 
at the weak scale $\mu=m_Z$.
Therefore, we must investigate 
how the mass ratios and CKM matrix parameters change 
from $\mu=\Lambda_X$ down to $\mu=m_Z$.
 \cite{evol}
In this paper, we distinguish between the values at $\mu=\Lambda_X$
and $\mu=m_Z$ by using the superscript "0" or not.

\section{Numerical study}
Because $M_u^0$, $M_d^0$, and $M_e^0$ are symmetric at the unification scale 
$\mu=\Lambda_X$
in the model with one 10 and one 126 Higgs scalars, 
they are diagonalized by unitary matrices $U_u^0$, $U_d^0$, and $U_e^0$, 
respectively, as 
\begin{equation}
U_u^{0\dagger}M_u^0 U_u^{0\star}=D_u^0 \ , \ \
U_d^{0\dagger}M_d^0 U_d^{0\star}=D_d^0 \ \mbox{\ and  \ }
U_e^{0\dagger}M_e^0 U_e^{0\star}=D_e^0 \ , \label{diag}
\end{equation}
where $D_u^0$, $D_d^0$, and $D_e^0$ are diagonal matrices which are given by
\begin{eqnarray}
D_u^0 &\equiv& \frac{1+\tan\beta^{-2}}{v^2} \mbox{diag}(m_u^0,m_c^0,m_t^0) \ , \ \
D_d^0 \equiv \frac{1+\tan\beta^{ 2}}{v^2} \mbox{diag}(m_d^0,m_s^0,m_b^0), \nonumber\\
D_e^0 &\equiv& \frac{1+\tan\beta^{ 2}}{v^2} \mbox{diag}(m_e^0,m_\mu^0,m_\tau^0) \ , 
\label{diag}
\end{eqnarray}
Here, \(v\)(\(=174\) GeV) is VEV of Higgs ,
and it is divided into up and down quark (neutrino and charged lepton) 
in the ratio \(\tan\beta\).
Using the Cabibbo-Kobayashi-Maskawa (CKM) matrix $V_q^0$ which is expressed as 
\(V_q^0=U_u^{0\dagger} U_d^0\), 
the relation (\ref{eq82515}) is rewritten as follows: 
\begin{equation}
(U_e^{0\dagger}U_u^0)^\dagger D_e^0(U_e^{0\dagger}U_u^0)^\star
= c_uD_u^0
+ c_{d}V_q^0 D_{d}V_q^{0 T}
= c_u(D_u^0
+ \kappa V_q^0 D_{d}V_q^{0 T}).\label{eq071703}
\end{equation}
We take a basis on which the up-quark Yukawa coupling is diagonal 
in order to compare with the experiment values 
and obtain the independent two equations:
\begin{eqnarray}
A \left( \kappa  \right) 
&\equiv& \frac{{\left( {\left( {m_e^0 m_\mu ^0 } \right)^2  + \left( {m_\mu ^0 m_\tau ^0 } \right)^2 
 + \left( {m_\tau ^0 m_e^0 } \right)^2 } \right)}}{{\left( {\left( {m_e^0 } \right)^2 
 + \left( {m_\mu ^0 } \right)^2  + \left( {m_\tau ^0 } \right)^2 } \right)^2 }}
   \frac{{2\left[ {{\mathop{\rm Tr}\nolimits} 
   \left\{ {H_q\left( \kappa  \right)} \right\}} \right]^2 }}
   {{\left\{ {{\mathop{\rm Tr}\nolimits} \left( {H_q\left( \kappa  \right)} \right)} \right\}^2 
 - {\mathop{\rm Tr}\nolimits} \left\{ {\left( {H_q\left( \kappa  \right)} \right)^2 } \right\}}}
 \to 1 \nonumber \\ 
B\left( \kappa  \right) &\equiv& \frac{{\left( {m_e^0 m_\mu ^0 m_\tau ^0 } \right)^2 }}
  {{\left( {\left( {m_e^0 } \right)^2  + \left( {m_\mu ^0 } \right)^2 
  + \left( {m_\tau ^0 } \right)^2 } \right)^3 }}
  \frac{{\left[ {{\mathop{\rm Tr}\nolimits} 
  \left\{ {H_q\left( \kappa  \right)} \right\}} \right]^3 }}
 {{\det \left\{ {H_q\left( \kappa  \right)} \right\}}} \to 1 
\label{eq1012-02}
\end{eqnarray}
Here \(H_q(\kappa)\) is the following hermite matrix which is defined by 
the Yukawa couplings of quark:
\begin{equation}
H_q(\kappa) \equiv (D_u^0+{\kappa}V^0D_{d}^0V^{0\dagger})
(D_u^0+{\kappa}V^0D_{d}^0V^{0\dagger})^\dagger .
\end{equation}
If we find the \(\kappa\) which sets \(A(\kappa)\) and \(B(\kappa)\) to \(1\) 
simultaneously,
the three Yukawa couplings \(M_u^0\), \(M_d^0\) and \(M_e^0\) can 
be unified into two matrices.
However we don't know precisely how to determine these data, especially quark masses.
And above procedures depend on these ambiguities.
So in this paper, we substitute the random numbers which becomes 
following normal distributions \cite{expmass}:
\begin{eqnarray}
 \left| {m_u \left( {2{\rm{GeV}}} \right)} \right| &=& 2.9 \pm 0.6{\rm{MeV}}, \quad
 \left| {m_d \left( {2{\rm{GeV}}} \right)} \right| = 5.2 \pm 0.9{\rm{MeV}}, \label{eq011901} \\
 \left| {m_s \left( {2{\rm{GeV}}} \right)} \right| &=& 80 - 155{\rm{MeV}}, \quad
 \left| {m_c \left( {m_c } \right)} \right| = 1.0 - 1.4{\rm{GeV}},\\
 m_b \left( {m_b } \right) &=& 4.0 - 4.5{\rm{GeV}}, \quad
 m_t^{{\rm{direct}}}  = 174.3 \pm 5.1{\rm{GeV}},
\end{eqnarray}
\begin{eqnarray}
 \left| {m_e^{{\rm{pole}}} } \right| &=& {\rm{0.510998902}} \pm {\rm{0.000000021MeV}},\\ 
 \left| {m_\mu ^{{\rm{pole}}} } \right| &=& {\rm{105.658357}} \pm {\rm{0.00005,}} \quad
 m_\tau ^{{\rm{pole}}} = {\rm{1776.99}} \pm {\rm{0.29MeV}} 
\end{eqnarray}
\begin{eqnarray} 
 \sin \theta _{12} &=& 0.2229 \pm 0.0022, \quad
 \sin \theta _{23} = 0.0412 \pm 0.0020,\\
 \sin \theta _{13} &=& 0.0036 \pm 0.0007, \quad
 \delta  = \left( {59 \pm 13} \right)^\circ \label{eq011902}
\end{eqnarray}
for each mass and CKM mixing parameter 10,000 times.
And we estimate the evolution effect about the values 
in Eqs. (\ref{eq011901}) - (\ref{eq011902}) 
from \(\mu=m_Z\) to \(\mu=\Lambda_X\) by using of RGE.\cite{evol}
In this work, we suppose MSSM for \(\mbox{tan}\beta=10\).
Without loss of generality, we can make the masses of third generation positive 
real number.
Although the remaining  masses are complex under the ordinary circumstances, 
we assume that all masses are real in order to simplify the problem.
Therefore, there are 16 combinations of the signs of the masses as shown 
in table \ref{tab1}.

As shown in Fig.1, we scan the range \(A(\kappa)=1\) by changing \(\mbox{Im}(\kappa)\) 
from -100 to 100 at 2000 equal intervals.
Moreover, we get the maximum and minimum of \(B(\kappa)\) on the line of \(A(\kappa)=1\) 
by changing \(\mbox{Im}(\kappa)\) at 5000 equal intervals.
Because \(B(\kappa)\) is continuous, 
there is the \(\kappa\) which sets \(A(\kappa)\) and \(B(\kappa)\) to \(1\) 
simultaneously when \(\mbox{Min}(B(\kappa))<1<\mbox{Max}(B(\kappa))\) 
as explained in Fig.2.
In this way, we draw the histograms in Figs 3,4,5 and 6 which show 
the distribution of input values conforming to the requirements 
\(\mbox{Min}(B(\kappa))<1<\mbox{Max}(B(\kappa))\).
The each summation of the conforming case is tabulated in Table 2 
after the 10,000 substitutions.
Expressed in another way, Table 2 shows the number of dots in the white area in Fig.2.
In Fig. 7, each circle in the complex plane shows the value of \(\kappa\) to 
meet the requirement \(A(\kappa)=B(\kappa)=1\), and
the total number of circle in each figure corresponds to 
the number in Table 2, obviously. 
From these figures and tables, it is understandable that 
the sign of \(m_u\) is not important.
Perhaps the reason is that \(m_u\) is very small, 
and it is almost negligible in comparison with other masses. 

\section{conclusion and discussion}
In conclusion, we have discussed the probability that the following model will 
be realized without fine tuning. 
The random numbers which become normal distributions have been 
substituted for each physical value at \(\mu=m_Z\). 
And we have taken the RGE effect between \(\mu=m_Z\) and \(\Lambda_X\) 
into consideration.
In this way, the search for \(\kappa\) which sets \(A(\kappa)\) and \(B(\kappa)\) to 
1 simultaneously has been repeated 10,000 times.
By this way, we have arrived at three conclusions:
(1) The probability that the model will be realized without 
fine tuning is about \(5\%\) if we select the appropriate signs (14) or (15) 
of the masses.
(2) This probability will increase if the signs of \(m_d\), and \(m_s\) are same.
This gives the suggestion to the texture model.
For example, a model with a texture \((M_d)_{11}=0\) 
on the nearly diagonal basis of the up-quark Yukawa coupling \(M_u\) is denied 
because these model leads to \(m_d/m_s < 0\).
(3) From Fig.3-Fig.6, this probability will increase 
if we make \(m_s\) somewhat larger or smaller than 
the present experiment value properly.

In the present paper, we have demonstrate that 
the quark and charged lepton Yukawa coupling can be unified into only two matrices.
However, we have not referred to the neutrino masses and lepton flavor mixings.
The neutrino Yukawa coupling is given by 
\begin{eqnarray}
M_D^0 &=& \frac{{c_{e0} }}{{c_{d0} }}c_{u0} M_0^0  + \frac{{c_{e1} }}{{c_{d1} }}c_{u1} M_1^0, \\
M_\nu ^0  &=& c_{R0}^{ - 1} M_D^{0} M_1^{0 - 1} M_D^{0\, T}.
\end{eqnarray}
Concerning this problem, 
we have not been able to find the positive solutions within \(3\sigma\) 
which is written by the paper \cite{neutrino} for the present.
However, since there are many possibilities for the neutrino mass generation mechanism,
we are optimistic about this problem.

\section*{Acknowledgments}
The author is grateful to Y.Koide, H.Fusaoka, T.Fukuyama, T.Kikuchi and H.Nishiura 
for the useful comments. 
This work is supported by the JSPS Research Fellowships for Young Scientists, No.3700.

\begin{table}[htbp]
\begin{tabular}{l|ll||l|ll}
    & (\(m_u\), \(m_c\), \(m_t\)) & (\(m_d\), \(m_s\), \(m_b\)) &
    & (\(m_u\), \(m_c\), \(m_t\)) & (\(m_d\), \(m_s\), \(m_b\)) \\ \hline
(0) & \((+++)\) & \((+++)\) & (8) & \((+++)\) & \((+-+)\) \\ 
(1) & \((-++)\) & \((+++)\) & (9) & \((-++)\) & \((+-+)\) \\ 
(2) & \((+-+)\) & \((+++)\) & (10)& \((+-+)\) & \((+-+)\) \\ 
(3) & \((--+)\) & \((+++)\) & (11)& \((--+)\) & \((+-+)\) \\ 
(4) & \((+++)\) & \((-++)\) & (12)& \((+++)\) & \((--+)\) \\ 
(5) & \((-++)\) & \((-++)\) & (13)& \((-++)\) & \((--+)\) \\ 
(6) & \((+-+)\) & \((-++)\) & (14)& \((+-+)\) & \((--+)\) \\ 
(7) & \((--+)\) & \((-++)\) & (15)& \((--+)\) & \((--+)\) \\ 
\end{tabular}

\caption{The combinations of 
the signs of (\(m_u\), \(m_c\), \(m_t\)) and (\(m_d\), \(m_s\), \(m_b\)).
The signs of the charged lepton are negligible in Eq.(\ref{eq1012-02}).
}
\label{tab1}
\end{table}

\begin{table}[htbp]
\begin{tabular}{l|l||l|l||l|l||l|l}
    & sum &    & sum &     & sum &     & sum \\ \hline
(0) & 344 &(4) &  34 & (8) &  56 & (12)& 283 \\
(1) & 328 &(5) &  30 & (9) &  60 & (13)& 294 \\
(2) & 225 &(6) &  35 & (10)&  54 & (14)& 470 \\
(3) & 209 &(7) &  35 & (11)&  56 & (15)& 482 \\
\end{tabular}

\caption{The total number of the cases conforming to the requirements 
Min\((B(\kappa))\) \(<1<\) Max\((B(\kappa))\) after the 10,000 substitutions.}
\label{tab2}
\end{table}

\begin{figure}[htbp]
\begin{center}
\includegraphics[width=5cm]{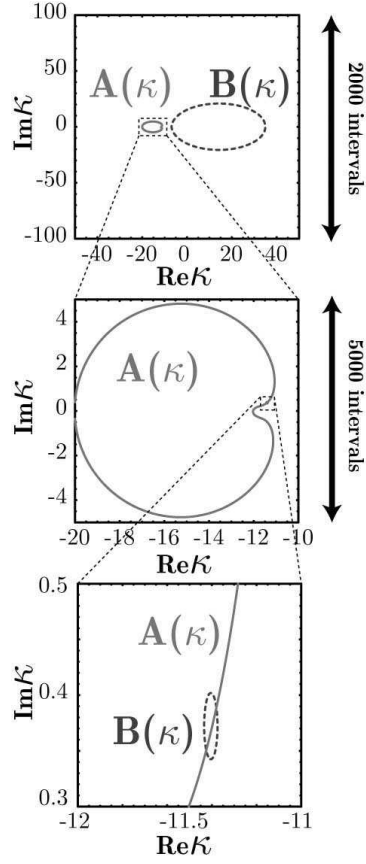}
\end{center}
\caption{The relations in Eq.(\ref{eq1012-02}) on the complex plane of \(\kappa\).
The solid line show \(A(\kappa)=1\) and the dotted line \(B(\kappa)=1\).
This is an example which is given as follows:
\(D_u^0\) = diag(\(-4.5116*10^{-6}\), \(-0.0011789\), \(0.5028  \)), 
\(D_d^0\) = diag(\(-5.6008*10^{-5}\), \(-0.0007776\), \(0.038776\)), 
\(D_e^0\) = diag(\( 1.8697*10^{-5}\), \( 0.0039461\), \(0.067375\)), 
\(\theta_{12}^0\) = \(0.22695\), \(\theta_{23}^0\) = \(0.035057\), 
\(\theta_{31}^0\) = \(0.0023936\), and
\(\delta^0\) = \(1.3173\).
}
\label{fig1}
\end{figure}

\begin{figure}[htbp]
\begin{center}
\includegraphics[width=7cm]{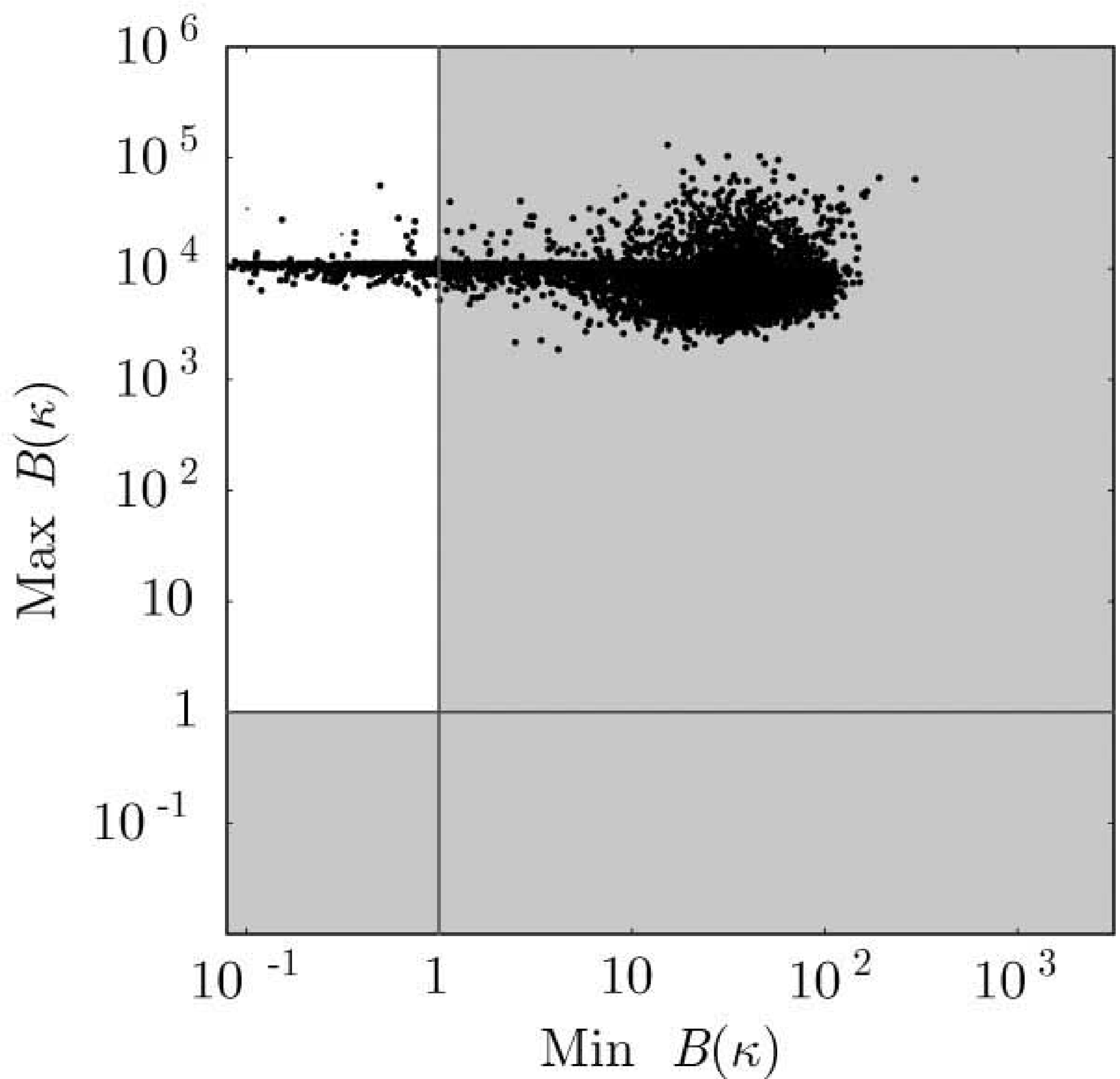}
\end{center}
\caption{The maximum and minimum of \(B(\kappa)\) on the line of \(A(\kappa)=1\) 
in the case of (15) in Table 1. There are 10000 dots in all area, 
and 482 dots in the white area as tabulated in Table 2.
}
\label{fig2}
\end{figure}

\begin{figure}[htbp]
\begin{center}
\includegraphics[width=16cm]{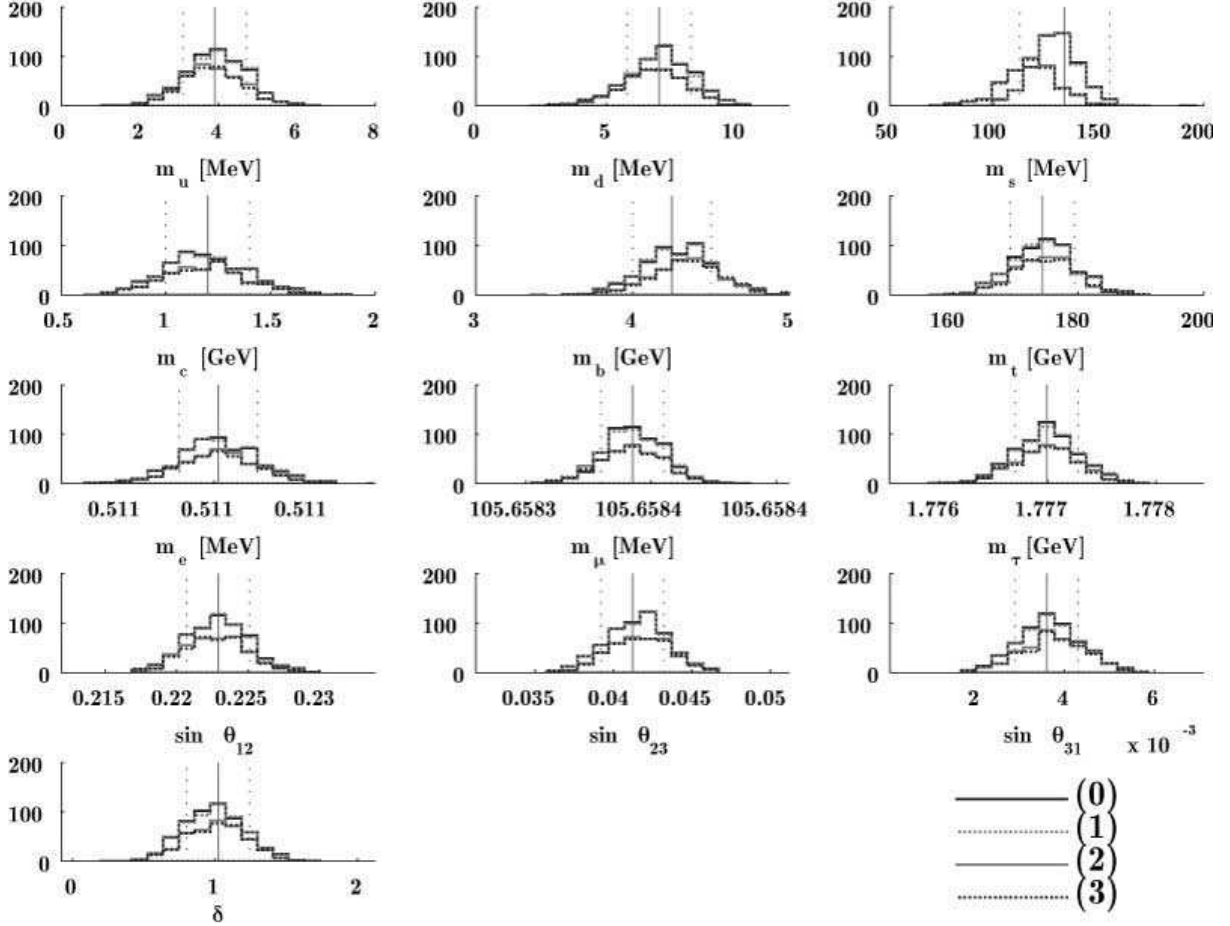}
\end{center}
\caption{The histograms show the distribution of data values 
conforming to the requirements \(\mbox{Min}(B(\kappa))<1<\mbox{Max}(B(\kappa))\).
Each number in parentheses show the signs of the mass in Table 1.
We bins the data values into 20 equally spaced containers, 
and show the number of elements in each container as a bar graph.
The vertical solid and dotted lines show the center value and range of error 
in Eqs.(\ref{eq011901}) - (\ref{eq011902}), respectively. 
}
\label{fig3}
\end{figure}

\begin{figure}[htbp]
\begin{center}
\includegraphics[width=16cm]{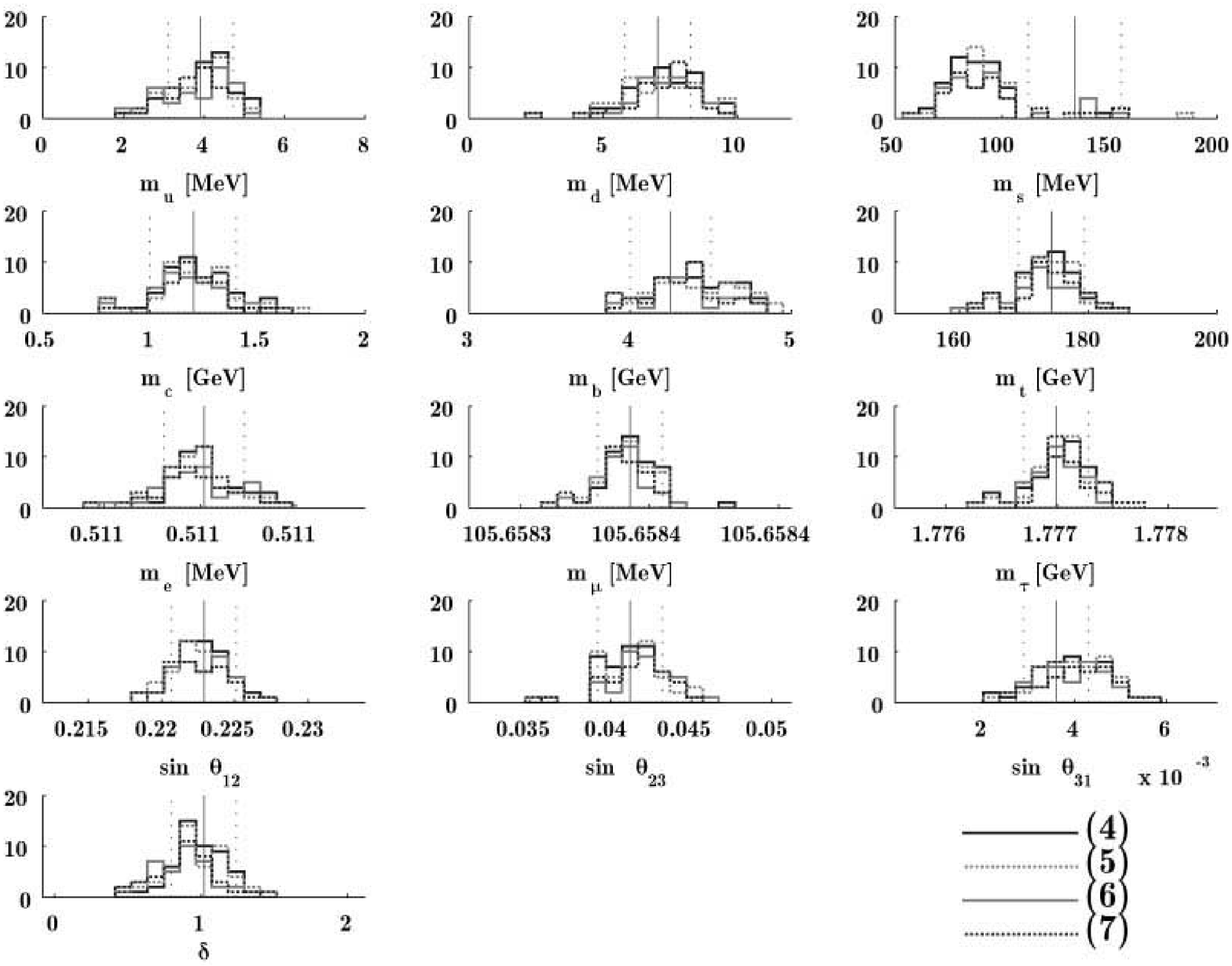}
\end{center}
\caption{The histograms show the distribution of data values as Fig.3.
}
\label{fig4}
\end{figure}

\begin{figure}[htbp]
\begin{center}
\includegraphics[width=16cm]{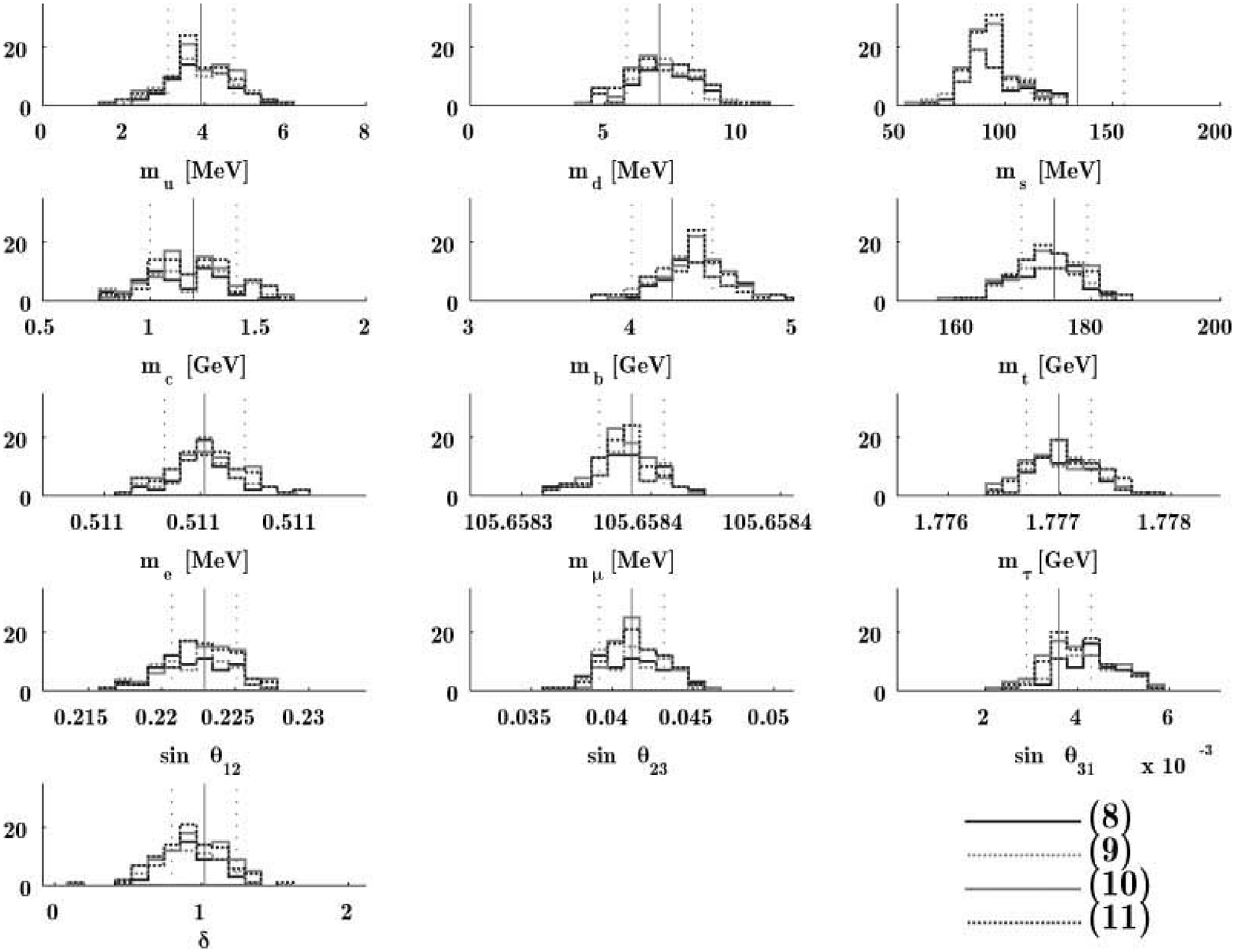}
\end{center}
\caption{The histograms show the distribution of data values as Fig.3.}
\label{fig5}
\end{figure}

\begin{figure}[htbp]
\begin{center}
\includegraphics[width=16cm]{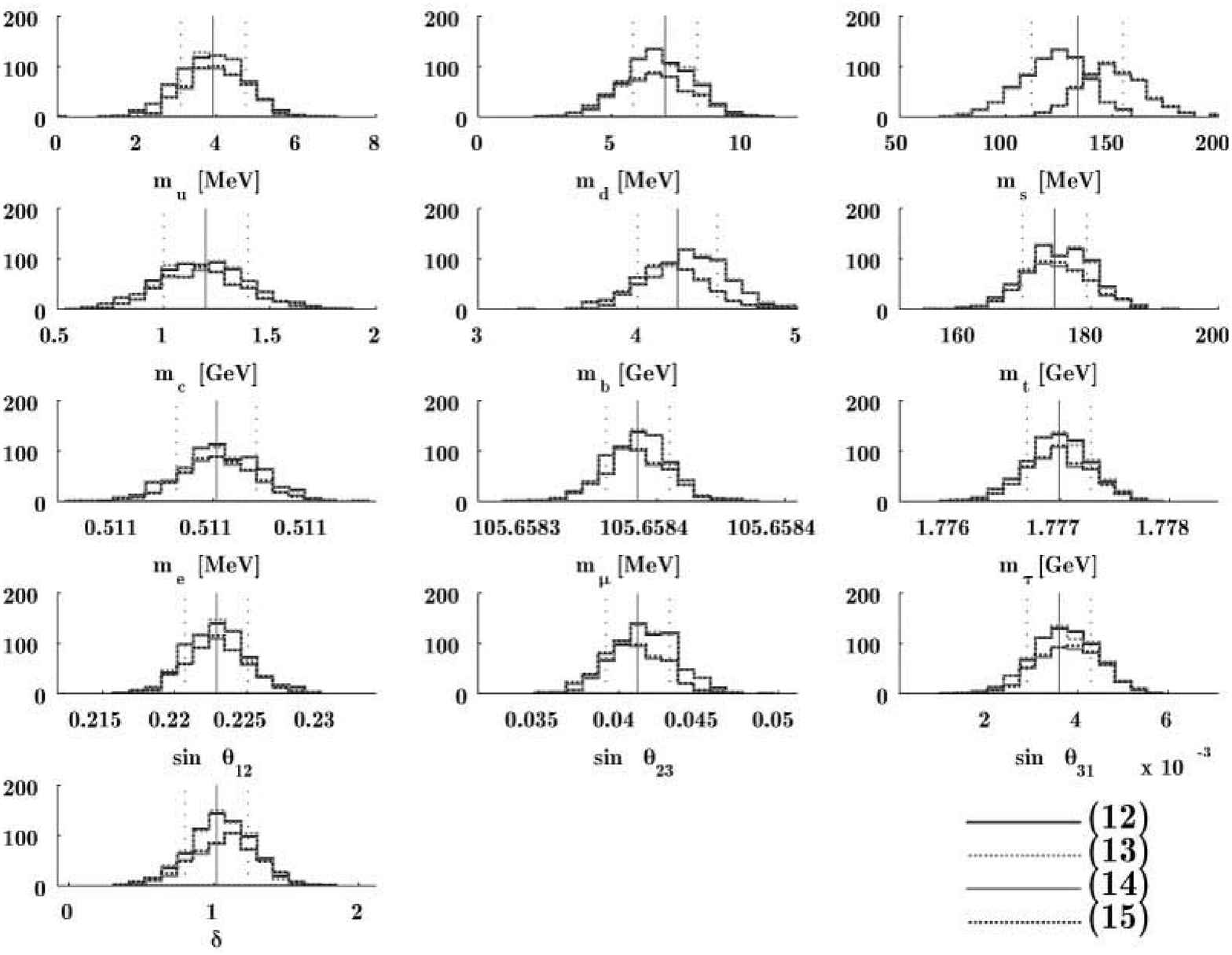}
\end{center}
\caption{The histograms show the distribution of data values as Fig.3.}
\label{fig6}
\end{figure}

\begin{figure}[htbp]
\begin{center}
\includegraphics[width=16cm]{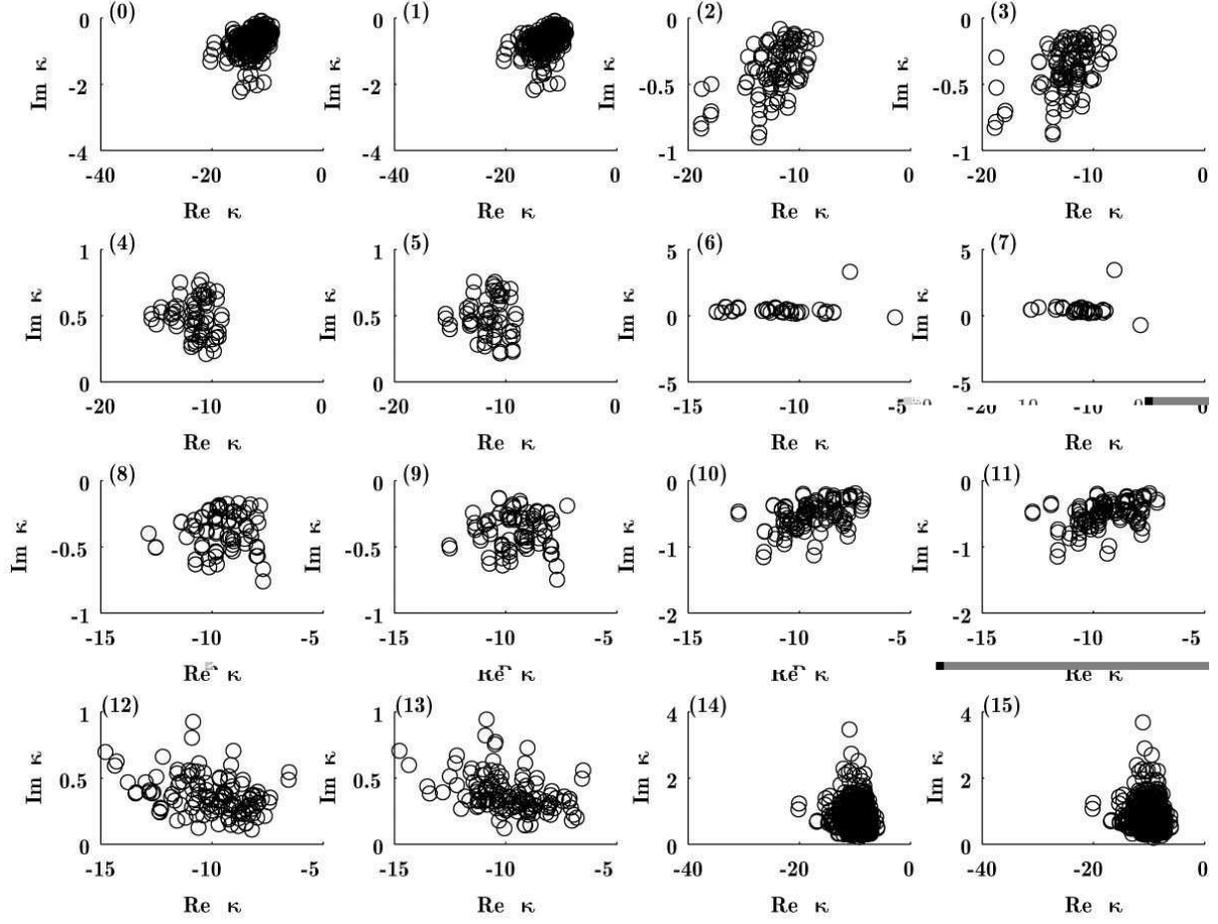}
\end{center}
\caption{The distribution of \(\kappa\) in the complex plane.
Each circle shows the value of \(\kappa\) to 
meet the requirement \(A(\kappa)=B(\kappa)=1\)}
\label{fig7}
\end{figure}

\end{document}